\documentclass[sigconf,10pt]{acmart}

\usepackage{graphicx}
\usepackage{amsmath}
\usepackage{amsfonts}
\usepackage{optidef}
\usepackage{multirow}
\usepackage{xcolor}
\usepackage{caption}
\usepackage{subcaption}
\usepackage{hyperref}
\usepackage{comment}
\usepackage{lipsum}
\usepackage{graphicx}
\usepackage[utf8]{inputenc}
\usepackage{enumitem}
\usepackage{hyperref}
\usepackage{amsmath}
\usepackage{amsfonts}
\usepackage{bm}
\usepackage{tabularx}
\usepackage{svg}
\usepackage{algpseudocode}
\usepackage{pgfplots}
\usepackage{tikz}
\usepackage{pgfplotstable}
\usepackage[utf8]{inputenc}
\pgfplotsset{compat=1.18}
\newlength\fheight
\newlength\fwidth
\usetikzlibrary{plotmarks,patterns,decorations.pathreplacing,backgrounds,calc,arrows,arrows.meta,spy,matrix}
\usepgfplotslibrary{patchplots,groupplots,colorbrewer}
\usepackage{tikzscale}
\usepackage{hyperref}
\newcommand{\WB}{\texttt{Wi-BFI}\xspace}

\newcommand{\mum}{\gls{mum}\xspace}

\newcommand{\bfi}{\gls{bfi}\xspace}

\newcommand{\bfas}{\glspl{bfa}\xspace}
\usepackage[linesnumbered,noend]{algorithm2e}

\AtBeginDocument{%
  \providecommand\BibTeX{{%
    \normalfont B\kern-0.5em{\scshape i\kern-0.25em b}\kern-0.8em\TeX}}}
\acmYear{2023}\copyrightyear{2023}
\setcopyright{acmlicensed}
\acmConference[ACM WiNTECH' 23]{The 17th ACM Workshop on Wireless Network Testbeds, Experimental evaluation \& Characterization 2023}{October 6, 2023}{Madrid, Spain}
\acmBooktitle{The 17th ACM Workshop on Wireless Network Testbeds, Experimental evaluation \& Characterization 2023 (ACM WiNTECH' 23), October 6, 2023, Madrid, Spain}
\acmPrice{15.00}
\acmDOI{10.1145/3615453.3616514}
\acmISBN{979-8-4007-0340-9/23/10}

\usepackage{xspace}
\SetKwRepeat{Do}{do}{while}

\usepackage[acronyms,nonumberlist,nopostdot,nomain,nogroupskip,hyperfirst=false]{glossaries}
\newacronym{iot}{IoT}{Internet of Things}
\newacronym{lpwan}{LPWAN}{low-power wide-area network}
\newacronym{sf}{SF}{Spreading Factor}
\newacronym{ml}{ML}{machine learning}
\newacronym{ai}{AI}{Artificial Intelligence}
\newacronym{nbiot}{NB-IoT}{Narrowband IoT}
\newacronym{fl}{FL}{Federated Learning}
\newacronym{arq}{ARQ}{Automatic Repeat Request}
\newacronym{fec}{FEC}{Forward Error Correction}
\newacronym{minlp}{MINLP}{Mixed-Integer Nonlinear Programming}
\newacronym{srn}{SRN}{Standard Radio Node}
\newacronym{sdr}{SDR}{Software Defined Radio}
\newacronym{cfr}{CFR}{channel frequency response}
\newacronym{mchem}{MCHEM}{Massive Channel Emulator}

\newacronym{rf}{RF}{radio frequency}
\newacronym{dsa}{DSA}{dynamic spectrum access}
\newacronym{rfp}{RFP}{radio fingerprinting}
\newacronym{shd}{SHD}{spectrum hole detection}
\newacronym{fml}{FML}{federated machine learning}
\newacronym{5g}{5G}{fifth generation}
\newacronym{mmw}{mmWave}{millimeter wave}
\newacronym{pus}{PUs}{primary users}
\newacronym{sus}{SUs}{secondary users}
\newacronym{drl}{DRL}{deep reinforcement learning}
\newacronym{dl}{DL}{deep learning}
\newacronym{dnn}{DNN}{deep neural networks}
\newacronym{ism}{ISM}{Industrial, Scientific, and Medical}
\newacronym{phy}{PHY}{physical layer}
\newacronym{css}{CSS}{chirp spread spectrum}
\newacronym{crc}{CRC}{cyclic redundancy check}

\newacronym{ap}{AP}{access point}
\newacronym{cnn}{CNN}{convolutional neural network}
\newacronym{csi}{CSI}{channel state information}
\newacronym{cv}{CV}{computer vision}
\newacronym{har}{HAR}{human activity recognition}
\newacronym{lan}{LAN}{local-area network}
\newacronym{lstm}{LSTM}{long short-term memory}
\newacronym{mimo}{MIMO}{multiple-input multiple-output}
\newacronym{nic}{NIC}{network interface card}
\newacronym{ofdm}{OFDM}{orthogonal frequency-division multiplexing}
\newacronym{ofdma}{OFDMA}{orthogonal frequency-division multiple access}
\newacronym{siso}{SISO}{single-input single-output}
\newacronym{sta}{STA}{station}
\newacronym{wlan}{WLAN}{wireless local-area network}

\newacronym{bfi}{BFI}{beamforming feedback information}

\newacronym{bfa}{BFA}{beamforming feedback angle}

\newacronym{mum}{MU-MIMO}{multi-user multi-input multi-output}
\newacronym{sum}{SU-MIMO}{single-user multi-input multi-output}

\newacronym{fsl}{FSL}{Few-Shot Learning}

\newacronym{snr}{SNR}{signal-to-noise ratio}
\newacronym{ndp}{NDP}{null data packet}
\newacronym{svd}{SVD}{singular value decomposition}

\newacronym[plural=\gls{ltf}s,firstplural=long training fields (LTFs)]{ltf}{LTF}{long training field}

\RestyleAlgo{ruled}

\usepackage[many]{tcolorbox}
\usepackage{lipsum}

\definecolor{titlebg}{RGB}{100,22,72}
\definecolor{introbg}{RGB}{0,128,128}

\newtcolorbox{usecase}[1][]{
  breakable,
  enhanced,
  arc=0pt,
  outer arc=0pt,
  colframe=titlebg,
  colback=titlebg!05,
  overlay unbroken and first={
    \node[
      draw=titlebg,
      fill=titlebg,
      rotate=0,
      anchor=north west,
      text=white,
      font=\bfseries
    ]
    at (frame.north west)  
    {#1};
  }
}

\newtcolorbox{mission}[1][]{
  breakable,
  enhanced,
  arc=0pt,
  outer arc=0pt,
  colframe=introbg,
  colback=introbg!05,
  overlay unbroken and first={
    \node[
      draw=introbg,
      fill=introbg,
      rotate=0,
      anchor=north west,
      text=white,
      font=\bfseries
    ]
    at (frame.north west)  
    {#1};
  }
}

\usepackage{fancyhdr}

\begin{document}

\renewcommand{\headrulewidth}{0pt}

\fancyhf{}
\fancyhead[C]{To be presented at ACM WiNTECH, Madrid, Spain, October 6, 2023}

\thispagestyle{fancy}
\title[\texttt{Wi-BFI}: Extracting the IEEE 802.11 Beamforming Feedback ]{\texttt{Wi-BFI}: Extracting the IEEE 802.11 Beamforming Feedback Information from Commercial Wi-Fi Devices}

\author[K. F. Haque, F. Meneghello and F. Restuccia]{Khandaker Foysal Haque$^\dagger$, Francesca Meneghello$^*$ and Francesco Restuccia$^\dagger$ \vspace{0.1cm}}
\affiliation{%
\institution{$\dagger$ Institute for the Wireless Internet of Things, Northeastern University, United States}
\country{}
}
\affiliation{%
\institution{$*$ Department of Information Engineering, University of Padova, Italy}
\country{}
}


\begin{abstract}

Recently, researchers have shown that the \bfas used for Wi-Fi \gls{mimo} operations can be effectively leveraged as a proxy of the \gls{cfr} for different purposes. Examples are passive human activity recognition and device fingerprinting.
However, even though the \bfas report frames are sent in clear text, there is not yet a unified open-source tool to extract and decode the \bfas from the frames. To fill this gap, we developed \WB, the first tool that allows retrieving  Wi-Fi \bfas and reconstructing the \gls{bfi}~-- a compressed representation of the \gls{cfr}~-- from the \bfas frames captured over the air. The tool supports \bfas extraction within both IEEE 802.11ac and 802.11ax networks operating on radio channels with 160/80/40/20~MHz bandwidth. Both multi-user and single-user \gls{mimo} feedback can be decoded through \WB. The tool supports real-time and offline extraction and storage of \bfas and \gls{bfi}. The real-time mode also includes a visual representation of the channel state that continuously updates based on the collected data. \textbf{\WB code is open source and the tool is also available as a pip package\footnote{https://github.com/kfoysalhaque/Wi-BFI}.}\vspace{-0.2cm}
\end{abstract}



\thispagestyle{fancy} 
\maketitle
\vspace{-0.2cm}
\section{Introduction}

\glsresetall

Wi-Fi is the technology of choice for plug-and-play Internet connectivity in indoor environments such as homes, offices, shopping malls, and university campuses. Around $19.5$ billion of Wi-Fi devices are currently deployed worldwide and their number is expected to increase with around $3.8$ billion of devices shipped annually~\cite{WiFiAlliance}.

To support the increasing demand for connectivity, modern Wi-Fi systems rely on \gls{mimo} technology that allows concurrently transmitting several data streams to multiple receivers (\textit{beamformees}). \Gls{mimo} leverages the \gls{cfr} estimate performed by the beamformees during channel sounding to obtain precoding parameters for each data stream at each transmitter antenna. The estimate is referred to as the \gls{bfi} and is transmitted by the beamformees to the transmitter (\textit{beamformer} in a compressed form -- the \bfas (see Section~\ref{sec:mu-mimo}).
Interestingly, \textit{the \bfas are transmitted in clear text} and, in turn, can be captured by any network analyzer, e.g., Wireshark, and decoded without the need for decryption. This characteristic makes the \textit{compressed \bfas an appealing proxy to the \gls{cfr}} to obtain information about the radio propagation channel. In turn, \bfas (and \gls{bfi}) can be used for a wide range of applications that span from network optimization to wireless sensing.

\begin{figure}[!ht]
	\centering
	\includegraphics[width=\linewidth]{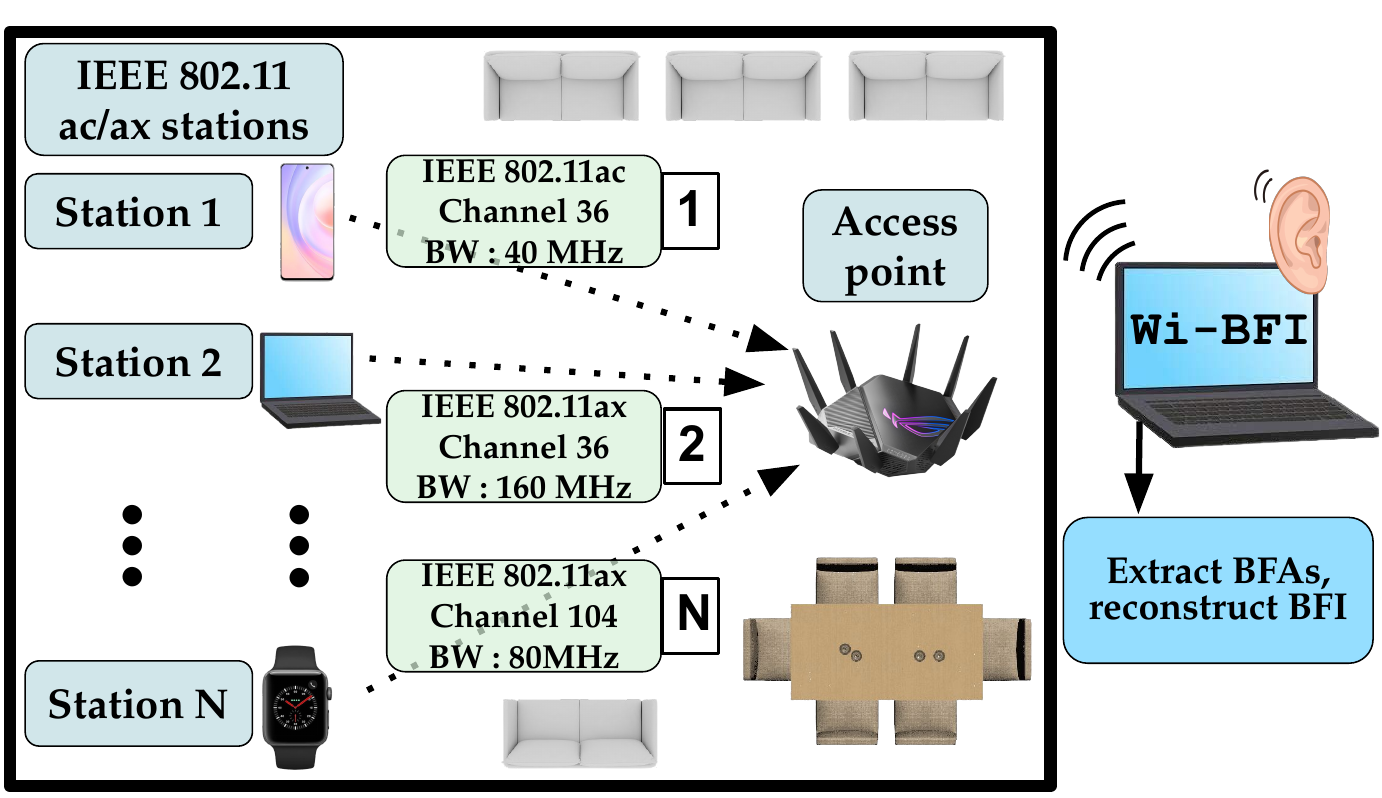}
 \setlength\abovecaptionskip{-0.3cm}
	\caption{\WB overview.\vspace{-0.6cm}}
	\label{fig:wi-bfi_overview}
\end{figure}

To enable this new paradigm, we developed \WB, the first tool that allows parsing the \bfas frames to extract the beamforming angles, and reconstruct the \gls{bfi}. An overview of \WB is depicted in Figure~\ref{fig:wi-bfi_overview}. \WB can concurrently extract the \bfas transmitted by devices implementing different Wi-Fi standards, and operating on channels with different bandwidths. For example, \WB can simultaneously extract the \bfas of an IEEE 802.11ac compliant device at 40~MHz (station 1 in Figure~\ref{fig:wi-bfi_overview}), and an IEEE 802.11ax compliant device at 160~MHz (station 2 in Figure~\ref{fig:wi-bfi_overview}) through a single capture. \WB does not require access to the monitored devices in the \gls{mimo} network. Contrarily, extracting the \gls{cfr} requires physical access and firmware modifications of the monitored devices~\cite{axcsi2021}. Because of this, state-of-the-art work has shown significant interest in \bfi~\cite{9700721, kondo2022bi}. Recent work has demonstrated that \bfas/\gls{bfi} can be successfully used in various applications like wireless human sensing~\cite{haque2023beamsense} and radio-fingerprinting~\cite{meneghello2022deepcsi}. While in previous work the procedure to extract the \bfas/\gls{bfi} was customized to the specific network setup, \textbf{\WB can be used with any network configuration} thus enabling further research in the area.\vspace{-0.25cm}

\subsection*{Summary of Paper Contributions}

This work provides the following contributions.\smallskip

\noindent$\bullet$ We propose \WB, an open source python-based tool to extract 802.11ac/ax \bfas in the wild and reconstruct the \gls{bfi}. \WB is compatible with any network configurations and operating systems.\smallskip

\noindent$\bullet$ We enable the extraction of \bfas and reconstruction of \gls{bfi} from multiple devices simultaneously, without any physical access to them. The monitored devices can implement different standards and operate on different bands.\smallskip 

\noindent$\bullet$ We provide both real-time plotting and saving of the extracted \bfas and reconstructed \gls{bfi} from live capture or any prior collected traces.\smallskip

\noindent$\bullet$ We present a use case entailing human activity classification with \bfas collected from IEEE 802.11ac devices operating at 80~MHz. By leveraging spatial diversity, the \bfas-based classifier achieves up to 99.28\% of accuracy.

\thispagestyle{fancy} 
\section{System Architecture of \texttt{W\lowercase{i}-BFI}}

In the following, we will use the superscripts $T$ and $\dag$ to denote the transpose and the complex conjugate transpose (i.e., the Hermitian). We define with $\angle{\mathbf{C}}$ the matrix containing the phases of complex-valued matrix $\mathbf{C}$. Moreover, diag$(c_1, \dots, c_j)$ indicates the diagonal matrix with elements $(c_1, \dots, c_j)$ on the main diagonal. The $(c_1, c_2)$ entry of matrix $\mathbf{C}$ is defined by $\left[\mathbf{\mathbf{C}}\right]_{c_1, c_2}$, while $\mathbb{I}_{c}$ refers to an identity matrix of size $c \times c$ and $\mathbb{I}_{c\times d}$ is a $c \times d$ generalized identity matrix.\vspace{-0.1cm}

\subsection{\WB Operation Principle}\label{sec:mu-mimo}

\WB leverages the way \gls{mimo} is implemented in Wi-Fi networks following the mechanism standardized in IEEE 802.11. To enable \gls{mimo}, the beamformer \textit{pre-codes} the data packets by linearly combining the signals to be simultaneously transmitted to the different beamformees. To do so, the beamformer uses a precoding matrix $\mathbf{W}$ which is derived from the \gls{cfr} matrix $\mathbf{H}$, describing how the environment modifies the irradiated signals in their path to the receiver. The \gls{cfr} needs to be estimated by each beamformee and fed back to the beamformer to allow proper pre-coding. 

\begin{figure}[h]
	\centering
	\includegraphics[width=.48\textwidth]{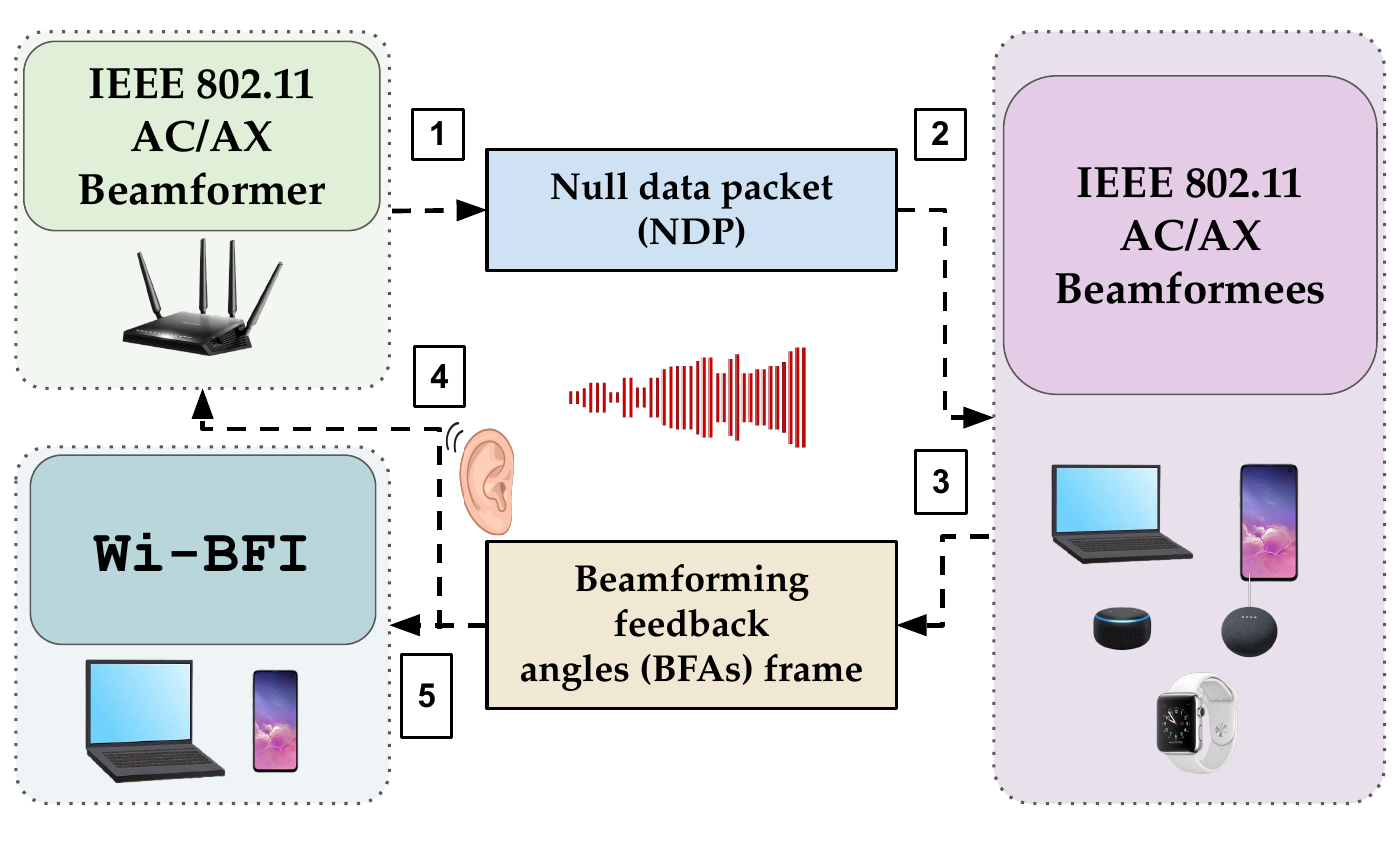}
  \setlength\abovecaptionskip{-0.4cm}
	\caption{\WB -- overall system architecture.\vspace{-0.2cm}}
	\label{fig:sounding_proc}
\end{figure}

The \gls{cfr} estimation process is called \textit{channel sounding} and is depicted in Figure~\ref{fig:sounding_proc}. The procedure is triggered by the beamformer that periodically broadcasts a \gls{ndp} to estimate the \gls{mimo} channel between itself and the connected beamformees (\textbf{step 1} in Figure~\ref{fig:sounding_proc}). Since its purpose is to sound the channel, the \gls{ndp} \textit{is not beamformed}. This is particularly advantageous as the resulting \textit{\gls{cfr} estimation is not affected by inter-stream or inter-user interference}.\vspace{-0.3cm}


The \gls{ndp} is a packet containing sequences of bits -- named \glspl{ltf} -- the decoded version of which is known by the beamformees. The \glspl{ltf} are transmitted over the different beamformer antennas in subsequent time slots, thus allowing each beamformee to estimate the \gls{cfr} of the links between its receiver antennas and the beamformer transmitter antennas. The \glspl{ltf} are modulated -- as the data fields -- through \gls{ofdm} by dividing the signal bandwidth into $K$ partially overlapping and orthogonal sub-channels spaced apart by $1/T$. The input bits are grouped into \gls{ofdm} symbols, \mbox{$\mathbf{a} = [a_{-K/2}, \dots, a_{K/2-1}]$} where $a_k$ is named \gls{ofdm} sample. The $K$ \gls{ofdm} samples are digitally modulated and transmitted through the $K$ \gls{ofdm} sub-channels in a parallel fashion thus occupying the channel for $T$ seconds. Thus, the transmitted \gls{ltf} signal is
\begin{equation}\label{eq:tx_signal}
	 s_{\rm tx}(t) = e^{j2\pi f_c t} \sum_{k=-K/2}^{K/2-1} a_{k} e^{j2\pi kt/T},
\end{equation}
where $f_c$ is the carrier frequency. 
The \gls{ndp} is received and decoded by each beamformee (\textbf{step 2} in Figure~\ref{fig:sounding_proc}) to estimate the \gls{cfr} $\mathbf{H}$. The different \glspl{ltf} are used to estimate the channel over each pair of transmitter (TX) and receiver (RX) antennas, for every \gls{ofdm} sub-channels. This generates a $K \times M \times N$ matrix $\mathbf{H}$ for each beamformee, where $M$ and $N$ are respectively the numbers of TX and RX antennas. Next, the \gls{cfr} is compressed -- to reduce the channel overhead -- and fed back to the beamformer. Using $\mathbf{H}_k$ to identify the $M \times N$ sub-matrix of $\mathbf{H}$ containing the \gls{cfr} samples related to sub-channel $k$, the \textit{compressed beamforming feedback} is obtained as follows (\cite{perahia_stacey_2008}, Chapter~13). First, $\mathbf{H}_k$ is decomposed through \gls{svd} as
\begin{equation}
    \mathbf{H}_k^T = \mathbf{U}_k\mathbf{S}_k\mathbf{Z}_k^\dag,
\end{equation}
where $\textbf{U}_k$ and $\mathbf{Z}_k$ are, respectively, $N \times N$ and $M \times M$ unitary matrices, while the singular values are collected in the $N\times M$ diagonal matrix $\mathbf{S}_k$. Using this decomposition, the complex-valued beamforming matrix $\mathbf{V}_k$ is defined by collecting the first $N_{\rm SS} \le N$ columns of $\mathbf{Z}_k$. Such a matrix is used by the beamformer to compute the pre-coding weights for the $N_{\rm SS}$ spatial streams directed to the beamformee. Hence, $\mathbf{V}_k$ is converted into polar coordinates to avoid transmitting the complete matrix. Specifically, $\mathbf{V}_k$ is decomposed as the product of $\mathbf{D}_{k,i}$ and $\mathbf{G}_{k,\ell,i}$ matrices, which are defined as
\begin{equation}
    \mathbf{D}_{k,i} =
	 \begin{bmatrix}
	\mathbb{I}_{i-1} & 0 & \multicolumn{2}{c}{\dots} & 0 \\
	0 & e^{j\phi_{k,i,i}} & 0 & \dots & \multirow{2}{*}{\vdots} \\
	\multirow{2}{*}{\vdots} & 0 & \ddots & 0 &  \\
	 & \vdots & 0 & e^{j\phi_{k,M-1,i}} & 0 \\
	0 & \multicolumn{2}{c}{\dots} & 0 & 1
	 \end{bmatrix},\label{eq:d_matrix}
\end{equation}
\begin{equation}
    \mathbf{G}_{k,\ell,i} =
	 \begin{bmatrix}
	\mathbb{I}_{i-1} & 0 & \multicolumn{2}{c}{\dots} & 0 \\
	0 & \cos{\psi_{k,\ell,i}} & 0 & \sin{\psi_{k,\ell,i}} & \multirow{2}{*}{\vdots} \\
	\multirow{2}{*}{\vdots} & 0 & \mathbb{I}_{\ell-i-1} & 0 &  \\
	 & -\sin{\psi_{k,\ell,i}} & 0 & \cos{\psi_{k,\ell,i}} & 0 \\
	0 & \multicolumn{2}{c}{\dots} & 0 & \mathbb{I}_{M-\ell}
	 \end{bmatrix},\label{eq:g_matrix}
\end{equation}
where the values of the $\phi$ and $\psi$ angles are obtained through the procedure detailed in Alg.~\ref{alg:beamf_feedback}. The number of $\phi$ and $\psi$ angles depends on the specific network configuration, i.e., the number of transmitter antennas and spatial streams.
Using these matrices and $\mathbf{\Tilde{D}}_k$ (see line 2 in Alg.~\ref{alg:beamf_feedback}), $\mathbf{V}_k$ can be written as $\mathbf{V}_k = \mathbf{\Tilde{V}}_k \mathbf{\Tilde{D}}_k$, with
\begin{equation}
    \mathbf{\Tilde{V}}_k = \prod_{i=1}^{\min(N_{\rm SS}, M-1)} \Bigg( \mathbf{D}_{k,i} \prod_{l=i+1}^{M}\mathbf{G}_{k,l,i}^T\Bigg)~ \mathbb{I}_{M\times N_{\rm SS}}, \label{eq:v_matrix}
\end{equation}
where the products represent matrix multiplications. In the $\mathbf{\Tilde{V}}_k$ matrix, the last row -- i.e., the feedback for the $M$-th transmitter antenna -- consists of non-negative real numbers by construction. Using this transformation, the beamformee is only required to transmit the $\phi$ and $\psi$ angles to the beamformer as they allow reconstructing $\mathbf{\Tilde{V}}_k$ precisely. Moreover, it has been proved (see \cite{perahia_stacey_2008}, Chapter~13) that the beamforming performance is equivalent at the beamformee when using $\mathbf{V}_k$ or $\mathbf{\Tilde{V}}_k$ to construct the steering matrix $\mathbf{W}$. In turn, the feedback for $\mathbf{\Tilde{D}}_k$ is not fed back to the beamformer. 

\RestyleAlgo{ruled}
\SetKwComment{Comment}{/*}{*/}
\SetAlgoNoLine
\LinesNotNumbered
\begin{algorithm}[t]
\caption{$\mathbf{V}_k$ matrix decomposition}\label{alg:beamf_feedback}
Require: $\mathbf{V}_k$\;
$\mathbf{\Tilde{D}}_k = {\rm diag}(e^{j \angle \left[\mathbf{V}_k\right]_{M,1}}, \dots, e^{j \angle \left[\mathbf{V}_k\right]_{M,N_{\rm SS}}})$ \;
$\mathbf{\Omega}_k = \mathbf{V}_k\mathbf{\Tilde{D}}_k^\dag$\;
\For{$i \leftarrow 1$ to $\min (N_{\rm SS}, M-1)$}{
$\phi_{k,\ell,i} = \angle \left[\mathbf{\Omega}_k\right]_{\ell, i}$ with $\ell={i, \dots, M-1}$\;
compute $\mathbf{D}_{k,i}$ through Eq.~(\ref{eq:d_matrix})\;
$\mathbf{\Omega}_k \leftarrow \mathbf{D}_{k,i}^\dag \mathbf{\Omega}_k$\;
\For{$\ell \leftarrow i+1$ to $M$}{
$\psi_{k,\ell,i} = \arccos \left( \frac{[\mathbf{\Omega}_k]_{i, i}}{\sqrt{[\mathbf{\Omega}_k]_{i, i}^2 + [\mathbf{\Omega}_k]_{\ell, i}^2}} \right)$\;
compute $\mathbf{G}_{k,\ell,i}$ through Eq.~(\ref{eq:g_matrix})\;
$\mathbf{\Omega}_k \leftarrow \mathbf{G}_{k,\ell,i} \mathbf{\Omega}_k$\;}}
\end{algorithm}
\setlength{\textfloatsep}{0.5cm}

\begin{figure*}
	\centering
	\includegraphics[width=0.72\textwidth, height=0.40\textwidth]{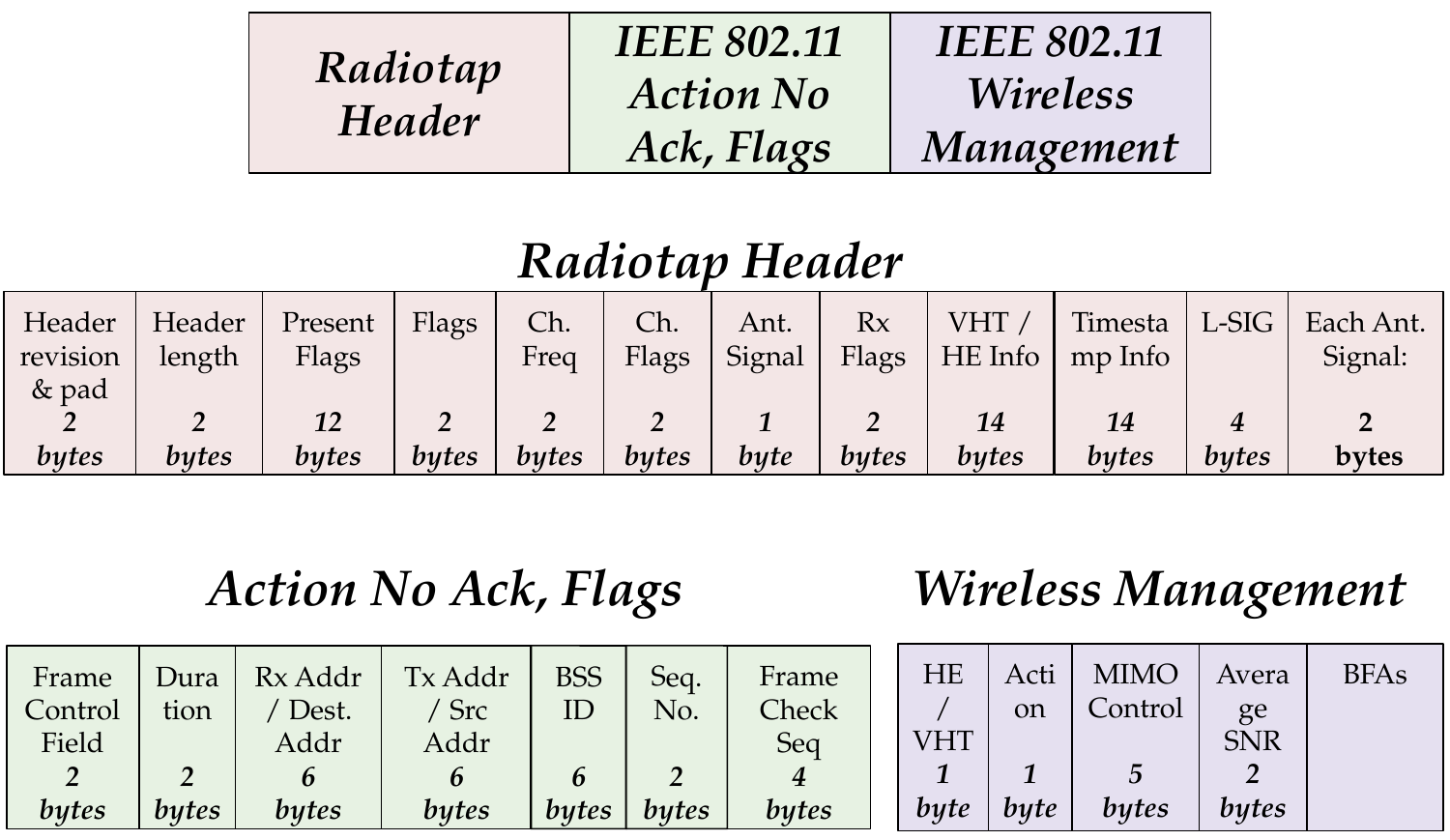}
	\caption{IEEE 802.11ac/ax BFI report frame structure.\vspace{-0.25cm}}
	\label{fig:bfi_frame}
\end{figure*}

To further reduce the channel occupancy, the angles are quantized using $b_{\phi}$ bits for $\phi$ and $b_{\psi} = b_{\phi}-2$ bits for $\psi$, as follows:
\begin{equation}
    [\phi, \psi] = \left[\pi \left ( \frac{1}{2^{b_\phi}} + 
    \frac{q_{\phi}}{2^{b_{\phi}-1}}\right),
    \left ( \frac{1}{2^{b_{\psi}+2}}+ \frac{q_{\psi}}{2^{b_{\psi}+1}} \right)\right] \label{eq:phi_psi}.
\end{equation}
In IEEE 802.11ac/ax, $b_{\phi}$ = \{9, 7\} bits and $b_{\phi}$ = \{6, 4\} bits are used for \gls{mum} and \gls{sum} systems respectively. The quantized values -- \mbox{$q_{\phi} = \{0, \dots, 2^{b_{\phi}}-1\}$} and \mbox{$q_{\psi} = \{0, \dots, 2^{b_{\psi}}-1\}$} -- are packed into the compressed beamforming frame (\textbf{step 3}  in Figure~\ref{fig:sounding_proc}) and such \textit{\bfas frame} is transmitted to the beamformer (\textbf{step 4} in Figure~\ref{fig:sounding_proc}). The $b_\phi$ and $b_\psi$ can be read in the VHT/HE \gls{mimo} control field, together with other information like the number of columns ($N_{\rm SS}$) and rows (M) in the beamforming matrix and the channel bandwidth (see Figure \ref{fig:bfi_frame}). Each \bfi contains $A$ angles for each of the $K$ \gls{ofdm} sub-channels for a total of $A \times K$ angles each. We remark that, since \mum requires fine-grained channel sounding -- around every 10 milliseconds, according to \cite{gast2013802} -- it is fundamental to process the \bfi in a fast manner at the beamformer. For this reason, and since cryptography would lead to excessive delays, \textit{the angles are currently transmitted over-the-air unencrypted}.

\WB captures the \bfas frames (\textbf{step 5}  in Figure~\ref{fig:sounding_proc}) with the python-based network monitoring tool \texttt{Pyshark}. However, any other network monitoring tool like \texttt{tcpdump} or \texttt{Wireshark} can also be used with \WB. 

Note that the beamformer continuously triggers the channel sounding procedure on the connected beamformees. This makes \textit{the \bfi contain very rich, reliable, and spatially diverse information}. Moreover, as \bfas frames are broadcasted by the beamformees, the information for all the beamformees \textit{can be collected with a single capture} by any Wi-Fi-compliant device. \WB leverages this procedure to collect the \bfas frames from multiple devices operating on different bands with a single capture, and reconstruct the \bfi from the extracted \bfas. Thus, \WB does not need any direct access to the beamformees or any hardware-specific tool, ultimately reducing the complexity of the data collection procedure.

\begin{figure}[t]
	\centering
	\includegraphics[width=.98\linewidth]{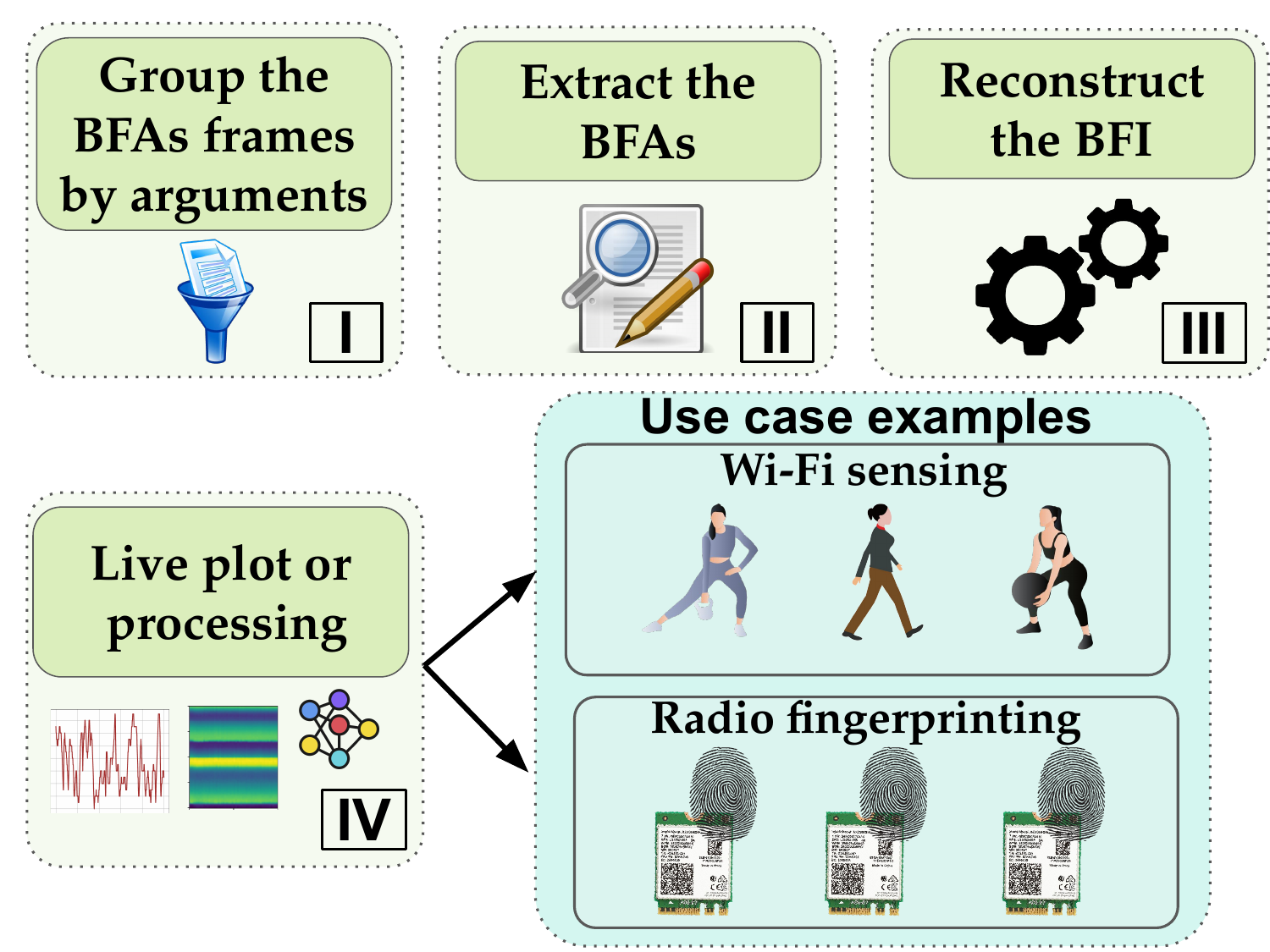}
        \setlength\abovecaptionskip{-0.001cm}
	\caption{Main steps of the \WB tool.\vspace{-0.4cm}}
	\label{fig:Wi-BFI_tool}
\end{figure}

\smallskip
\WB consists of four main steps as depicted in Figure~\ref{fig:Wi-BFI_tool}. Firstly, in \textbf{step I}, \WB groups the \bfas frames based on the following arguments: \textbf{(i)} the standard: IEEE 802.11ac or IEEE 802.11ax; \textbf{(ii)} the bandwidth: 20~MHz, 40~MHz, 80~MHz, or 160~MHz; \textbf{(iii)} the device, through the MAC address. 

\begin{figure}[t]
	\centering
	\includegraphics[width=.40\textwidth]{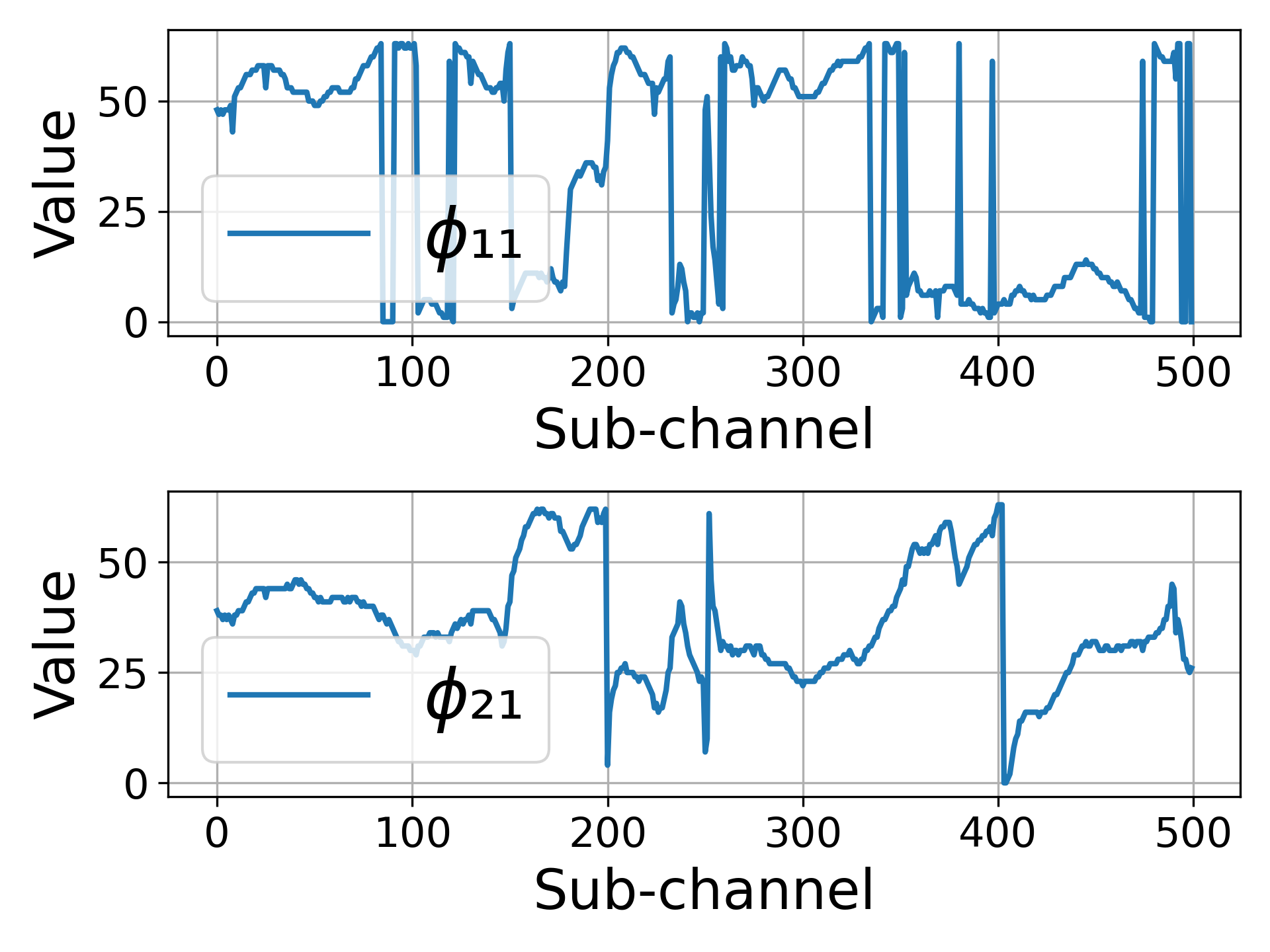}
        \setlength\abovecaptionskip{-0.05cm}
	\caption{\bfas($\phi_{11}$ and $\phi_{21}$) of each sounded sub-channel of a $\mathbf{4\times2}$ IEEE 802.11ax system at 160 MHz.\vspace{-0.05cm}}
	\label{fig:angles}
\end{figure}

After grouping the \bfas frames from any live capture or earlier captured trace, \WB performs bit-wise parsing through the frames to extract the \bfas (\textbf{step II} in Figure~\ref{fig:Wi-BFI_tool}). For that, we follow the IEEE 802.11 \bfas frame structure, presented in Figure \ref{fig:bfi_frame}. Specifically, a \bfas frame has three main parts: \textbf{(i)} \textit{Radiotap Header}, \textbf{(ii)} \textit{IEEE 802.11 Action No Ack}, \textit{Flags}, and \textbf{(iii)} \textit{IEEE 802.11 Wireless Management}. The fields of each part of a \bfas frame, along with the corresponding byte values, are mentioned in Figure~\ref{fig:bfi_frame}. \WB parses through each of the fields of the \bfas frame to extract the corresponding information including the \bfas. The \bfas are structured in a matrix of dimension is $ P\times K \times A$, where $P$ represents the total number of packets in the capture. As an example, considering a $4\times2$ IEEE 802.11ax \gls{sum} system operating on a 160~MHz bandwidth channel, with the beamformer having four antennas and the beamformee having two antennas, each \bfas frame has $K=$500 OFDM data sub-channels with $A=$10 \bfas each. For this setup, Figure \ref{fig:angles} depicts the first two \bfas ($\phi{_{11}}$ and $\phi{_{21}}$) for each of the OFDM sub-channels (x-axis) for one \bfas frame. 

Note that, according to the standard, the number of sub-channels changes with the change in operating standards and transmission bandwidth~\cite{IEEE802.11ax}. Moreover, we remind that the number of \bfas depends on the network configuration, i.e., the number of transmit antennas $M$, and the number of spatial streams $N_{\rm SS}$. The number and order of \bfas for some network configurations are reported in Table~\ref{table:angles}. 

\begin{table}[!ht]
\begin{tabular}{|p{1.2cm}|p{2.2cm}|p{3.7cm}|}
\hline
\normalsize
\textbf{MIMO} \centering & \textbf{No. angles} \centering & \textbf{Order of the angles}   \\ \hline

 $2\times1$ \centering    &   2  \centering    &   $\phi_{11}$, $\psi_{21}$      \\ \hline
 $2\times2$ \centering    &   2  \centering    &   $\phi_{11}$, $\psi_{21}$       \\ \hline
 $3\times1$ \centering    &   4  \centering    &   $\phi_{11}$, $\phi_{21}$, $\psi_{21}$, $\psi_{21}$                    \\ \hline
 $3\times2$ \centering    &   6  \centering    &   $\phi_{11}$, $\phi_{21}$, $\psi_{21}$, $\psi_{21}$, $\phi_{22}$, $\psi_{32}$  \\ \hline
 $3\times3$ \centering    &   6  \centering    &   $\phi_{11}$, $\phi_{21}$, $\psi_{21}$, $\psi_{21}$, $\phi_{22}$, $\psi_{32}$  \\ \hline
 $4\times1$ \centering    &   6  \centering    &   $\phi_{11}$, $\phi_{21}$, $\phi_{31}$, $\psi_{21}$, $\psi_{31}$, $\psi_{41}$  \\ \hline
 $4\times2$ \centering    &   10 \centering    &    \multicolumn{1}{l|}{\begin{tabular}[c]{@{}l@{}}$\phi_{11}$, $\phi_{21}$, $\phi_{31}$, $\psi_{21}$, $\psi_{31}$, $\psi_{41}$,\\$\phi_{22}$, $\phi_{32}$, $\psi_{32}$, $\psi_{42}$ \end{tabular}}             \\ \hline
 $4\times3$ \centering    &   12 \centering    &     \multicolumn{1}{l|}{\begin{tabular}[c]{@{}l@{}}$\phi_{11}$, $\phi_{21}$, $\phi_{31}$, $\psi_{21}$, $\psi_{31}$, $\psi_{41}$,\\$\phi_{22}$, $\phi_{32}$, $\psi_{32}$, $\psi_{42}$, $\phi_{33}$, $\psi_{43}$ \end{tabular}}         \\ \hline
 $4\times4$ \centering    &   12 \centering    &   \multicolumn{1}{l|}{\begin{tabular}[c]{@{}l@{}}$\phi_{11}$, $\phi_{21}$, $\phi_{31}$, $\psi_{21}$, $\psi_{31}$, $\psi_{41}$,\\$\phi_{22}$, $\phi_{32}$, $\psi_{32}$, $\psi_{42}$, $\phi_{33}$, $\psi_{43}$ \end{tabular}}       \\ \hline
\end{tabular}\vspace{0.25cm}
\caption {Order of \bfas in the BFI report frame in Figure~\ref{fig:bfi_frame} for different MIMO configurations.\vspace{-0.2cm}} \label{table:angles}
\end{table}


\begin{figure}[t]
	\centering
	\includegraphics[width=\linewidth]{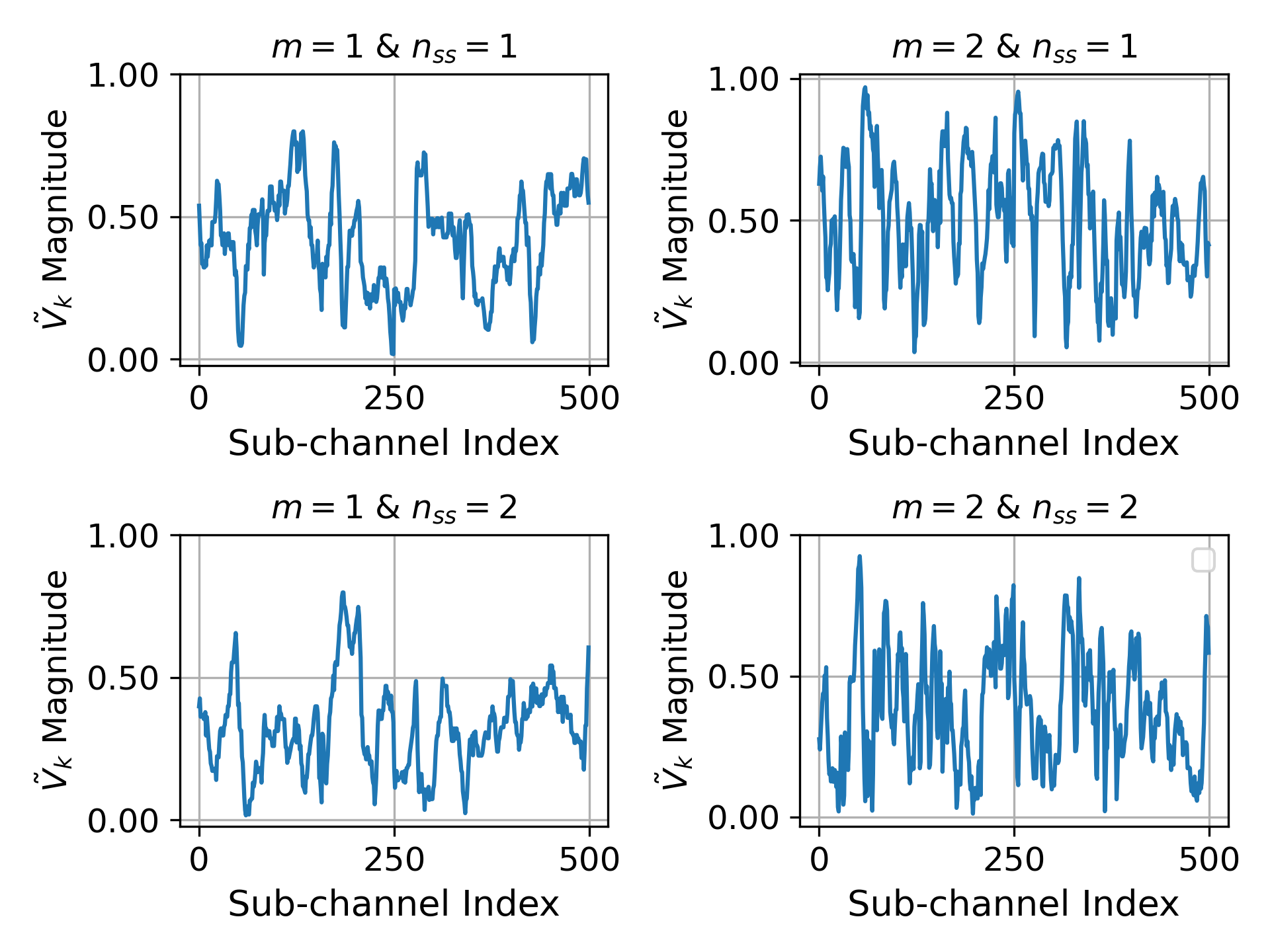}
        \setlength\abovecaptionskip{-0.2cm}
	\caption{Magnitude of $\mathbf{\tilde{V}_{k}}$ of each sounded sub-channel for different transmit antennas $m \in \{0, \dots, M-1\}$ and spatial streams $n_{\rm SS} \in \{0, \dots, N_{\rm SS}-1\}$ of a $\mathbf{4\times2}$ IEEE 802.11ax system at 160 MHz.\vspace{-0.5cm}}
	\label{fig:V_k}
\end{figure}

\begin{figure}[ht!]
	\centering
	\includegraphics[width=.40\textwidth]{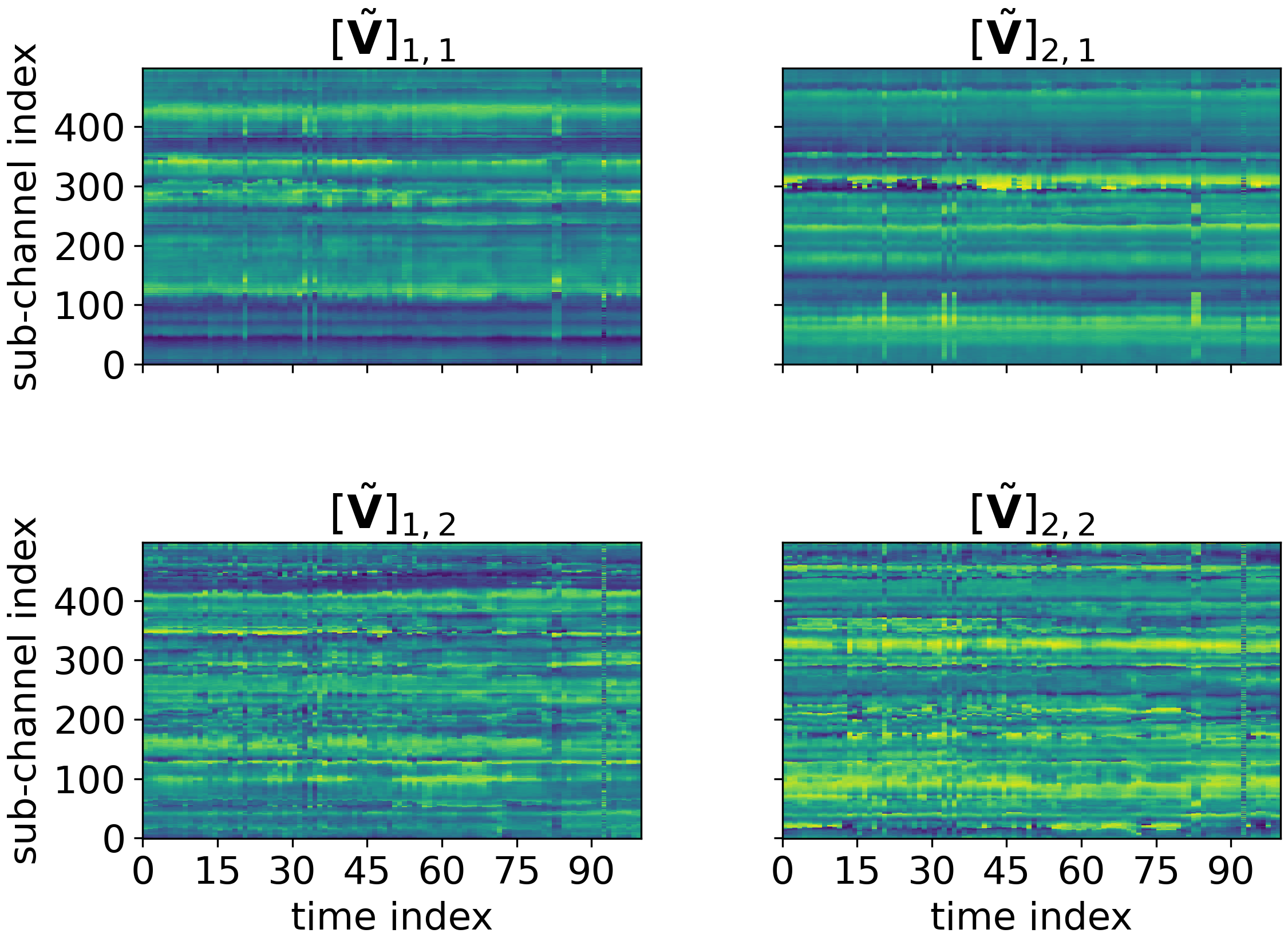}
        \setlength\abovecaptionskip{-0.01cm}
	\caption{Time evolution of $\tilde{\mathbf{V}}$. The columns refer to the transmit antenna and rows to the spatial streams.\vspace{-0.1cm}}
	\label{fig:V_spectrogram}
\end{figure}

After extracting the \bfas, \WB reconstructs the $\tilde{\mathbf{V}}_k$ matrix following Equation~\ref{eq:v_matrix} (see \textbf{step III} of Figure \ref{fig:Wi-BFI_tool}). The dimension of the reconstructed $\tilde{\mathbf{V}}_k$ matrix is $P \times K\times M \times N_{\rm SS}$. For the above mentioned $4\times2$ IEEE 802.11ax system at 160 MHz, the reconstructed $\tilde{\mathbf{V}}_k$ has $K$=500, $M$=4 and $N_{\rm SS}$= 2 making the dimension $P\times500\times4\times2$. The $\tilde{\mathbf{V}}_k$ matrices reconstructed by \WB from the captured angles for the first two transmitter antennas are depicted in Figure~\ref{fig:V_k}, i.e., for ($m=1$, $n_{\rm SS}=1$), ($m=1$, $n_{\rm SS}=2$), ($m=2$, $n_{\rm SS}=1$), and ($m=2$, $n_{\rm SS}=2$), with $m \in \{0, \dots, M-1\}$ and $n_{\rm SS} \in \{0, \dots, N_{\rm SS}-1\}$) being the transmitter antenna and the spatial stream indices, respectively. Figure~\ref{fig:V_k} reports $\tilde{\mathbf{V}}_k$ for a single frame. Figure~\ref{fig:V_spectrogram} depicts the time evolution of $\tilde{\mathbf{V}}_k$ (multiple frames). 

As the final step, as depicted in \textbf{step IV} of Figure \ref{fig:Wi-BFI_tool}, \WB either plots or saves the extracted \bfas and the $\mathbf{\tilde{V}_k}$ matrix (\bfi). For further visual representations of the \bfas, please refer to our \WB demonstration video.\footnote{\WB demonstration available at \url{https://youtu.be/0k7uYRCmMBw}}

\section{\WB Use case: Sensing with BFA\lowercase{s}}

Beyond connectivity, the unprecedented rise in the number of smartphones, laptops, and many other \mbox{Wi-Fi-enabled} devices~\cite{WiFiforIoT} will pave the way to revolutionary Wi-Fi sensing applications including activity recognition~\cite{haque2023beamsense}, and radio-fingerprinting~\cite{meneghello2022deepcsi}, among others~\cite{ma2019wifi}.

The vast majority of such applications leverage the uncompressed \gls{cfr} measurements obtained from pilot symbols in the Wi-Fi packets to characterize the propagation channel. Despite leading to good performance, \gls{cfr}-based techniques require manual extraction and recording of the \gls{cfr}, which is currently not supported by the IEEE 802.11 standard. This has led to the usage of \gls{cfr} extractor tools running custom-tailored firmware~\cite{csitool2011,axcsi2021}. Even if 802.11 could eventually support \gls{cfr} extraction in the future, (i) it would require additional device-specific processing to extract it from the chip, thus increasing energy consumption; (ii) multi-device sensing would require tight synchronization among collectors to align the time of the individual \gls{cfr} measurements. On top of these, most of the existing \gls{cfr} tools need direct access to the device and only work on devices implementing specific standards and operating on channels with certain bandwidths. This limits different applications to leverage the Wi-Fi signals opportunistically and hinders mass adoption at the same time.

On the contrary, we argue that the \gls{bfi}, and, in turn, the \bfas, are more effective metrics for developing sensing applications.
In this section, we show how \bfas can be leveraged for human activity classification through IEEE 802.11ac \gls{mum} systems operating on channels with 80 MHz of bandwidth. A comparison of \bfas and traditional \gls{cfr} based approach along with domain generalization capability of \bfas are presented in \cite{haque2023beamsense}.\vspace{-0.1cm}

\subsection{Learning Architecture}
\begin{figure}[ht]
	\centering
	\includegraphics[width=.96\linewidth]{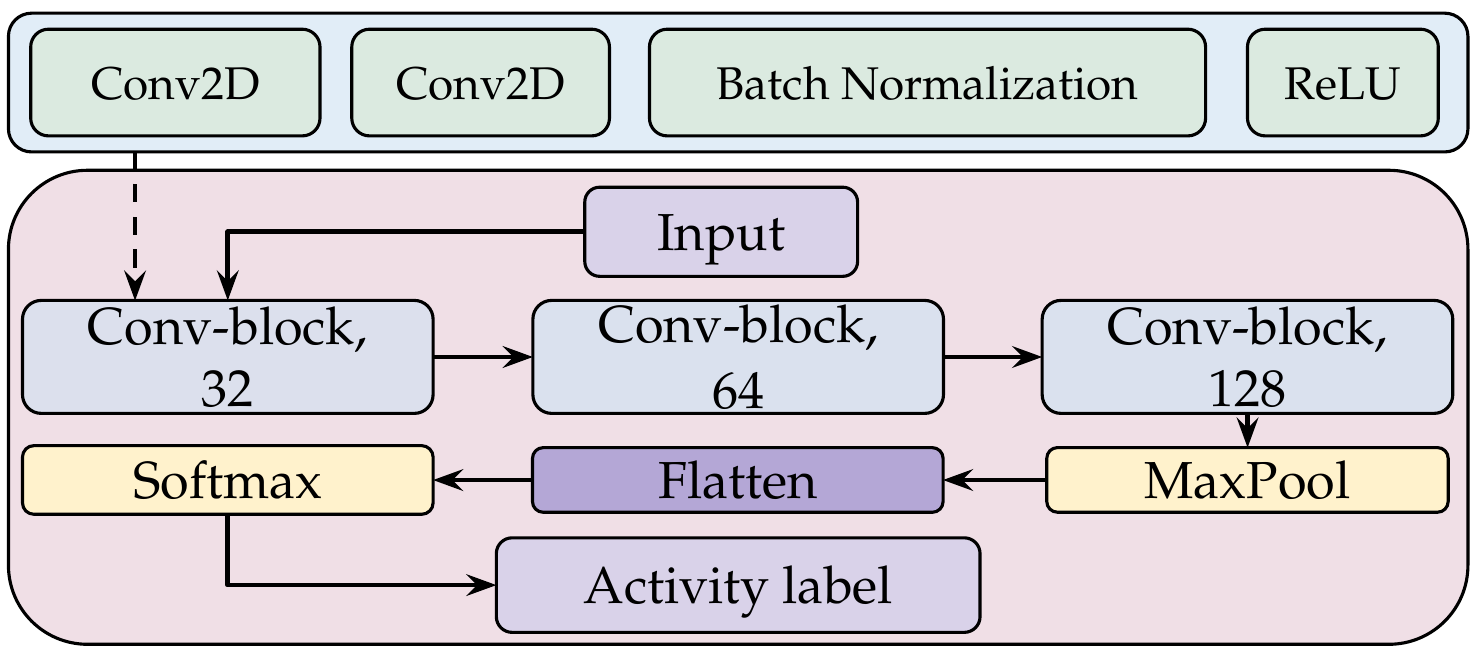}
	\caption{CNN-based learning architecture.\vspace{-0.4cm}}
	\label{fig:baseline_cnn}
\end{figure}

Owing to its excellent performance, \gls{cnn} is a popular choice in a wide range of applications including wireless sensing \cite{haque2023beamsense, haque2023simwisense}. The convolutional layer is the basis of \gls{cnn} which performs convolution operations on the elements of the input data to extract features. Thus, for demonstrating the potential of \bfas in Wi-Fi sensing tasks like activity classification, we leverage a \gls{cnn} architecture based on only three VGG-based blocks~\cite{simonyan2014very}. 
\noindent
Specifically, the network stacks three convolutional blocks (\texttt{conv-block}) and a max-pooling (\texttt{MaxPool}) layer as detailed in Figure~\ref{fig:baseline_cnn}. The output of the \texttt{MaxPool} layer is flattened and then the Softmax activation function is used to obtain the probability distribution over the activity labels.

Each \texttt{conv-block} consists of two 2-dimensional convolutional layers. Inspired by the design of VGG block, we use a kernel size of $3\times3$ and a step size of 1 in each of the convolutional layer~\cite{simonyan2014very}. We use batch normalization and rectified linear units (ReLU) activation function in each \texttt{conv-block} to avoid gradient explosion or vanishing and to have non-linearity in the network respectively. Our VGG-based \gls{cnn} consists of three of such \texttt{conv-blocks} with 32, 64, and 128 filters respectively.
\vspace{-0.15cm}

\subsection{Experimental Setup}

\begin{figure}[ht]
	\centering
	\includegraphics[width=.96\linewidth]{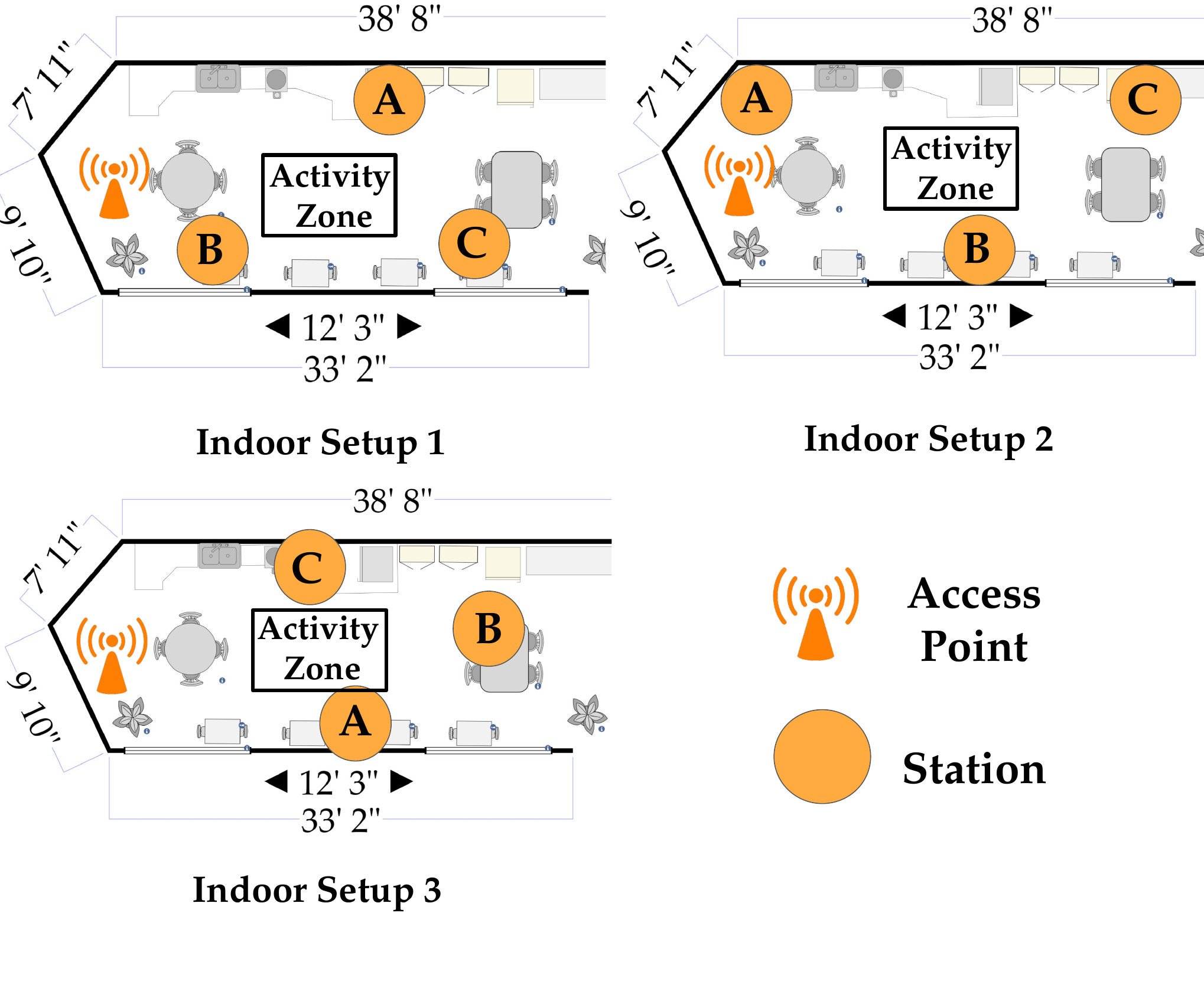}
        \setlength\abovecaptionskip{-0.5cm}
	\caption{Experimental setup for activity classification with \bfas.\vspace{-0.3cm}}
	\label{fig:sensing_setup}
\end{figure}

The experiments are conducted in three indoor scenarios with three different subjects as presented in Figure \ref{fig:sensing_setup}. 

\begin{figure}[!ht]
	\centering
	\includegraphics[width=.45
\textwidth]{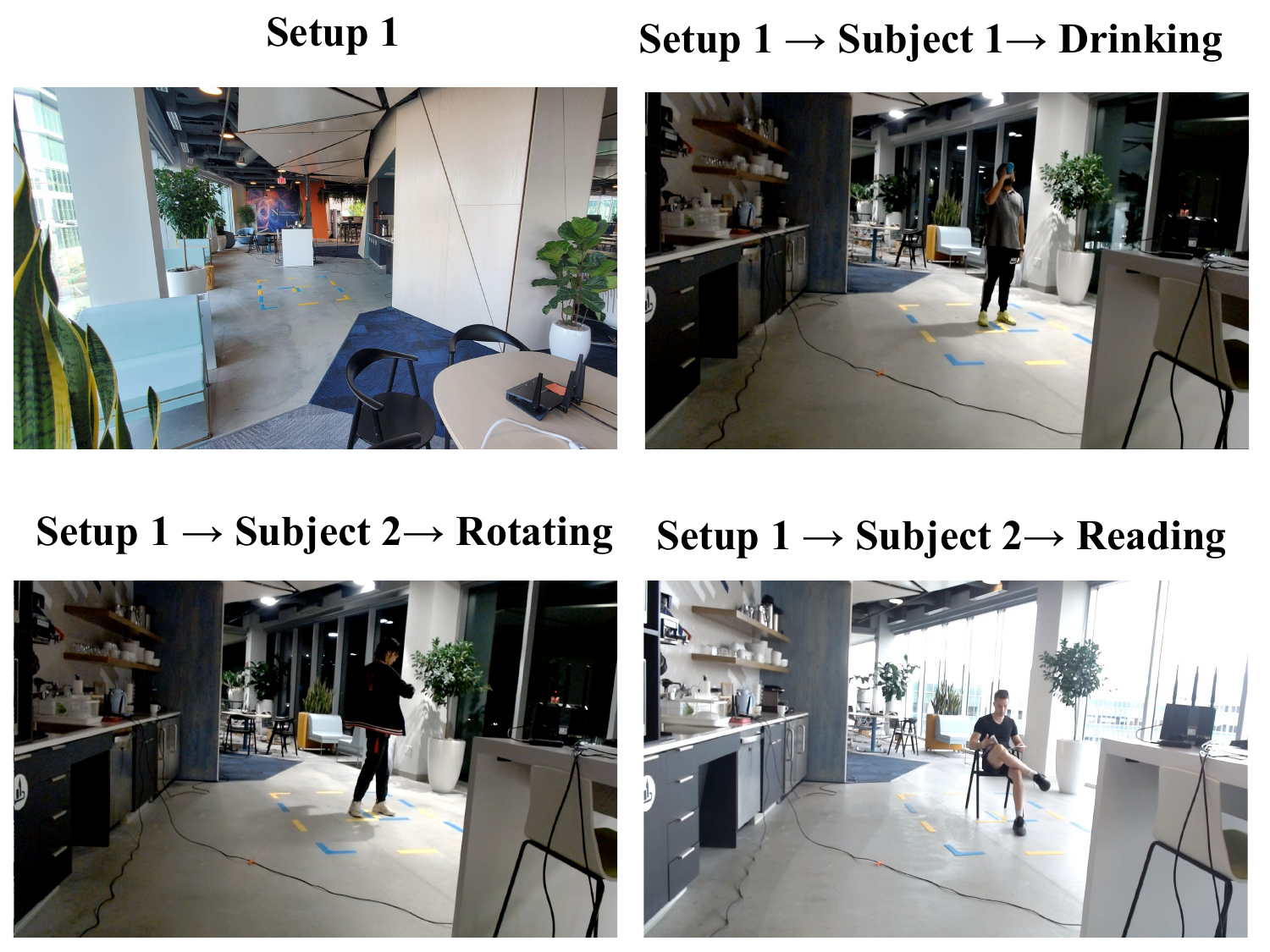}
	\caption{Experimental setup for activity classification with \bfas.\vspace{-0.45cm}}
	\label{fig:setup_capture}
\end{figure}

In each of the setups, we have a $3\times3$ IEEE 802.11ac \mum system operating on channel 153 with center frequency $f_c=5.77$ GHz, with 80~MHz of bandwidth. Each of the systems has one beamformer and three beamformees (\glspl{sta}) with M=3 and N=1 antennas enabled respectively. We use off-the-shelf Netgear Nighthawk X4S AC2600 routers both as beamformer and beamformees to set up the network. We consider three human subjects performing ten different activities: \textit{walking, standing, rotating, waiving, drinking, hands up and down, reading a book, jogging, boxing, and clapping.} UDP data streams are sent from the beamformer to the beamformees in the downlink direction to trigger the channel sounding. We record the \bfas frames of all three stations with a single capture with \WB and extract the \bfas associated with each beamformees. For training the model we capture the \bfas frames for 3 minutes for each activity performed by each subject. Hence, for each of the beamformees we divide the extracted \bfas into non-overlapping windows of 10 packets each. The windows are fed one at a time to the learning block. To obtain the ground truth, we also capture the video streams of the subjects performing the activities. The experimental setup and the snaps from the ground truth video streams with three different subjects and activities are shown in Figure~\ref{fig:setup_capture}.

\subsection{Sensing Performance with \bfas}

\begin{figure}[ht]
	\centering
	\includegraphics[width=.43\textwidth]{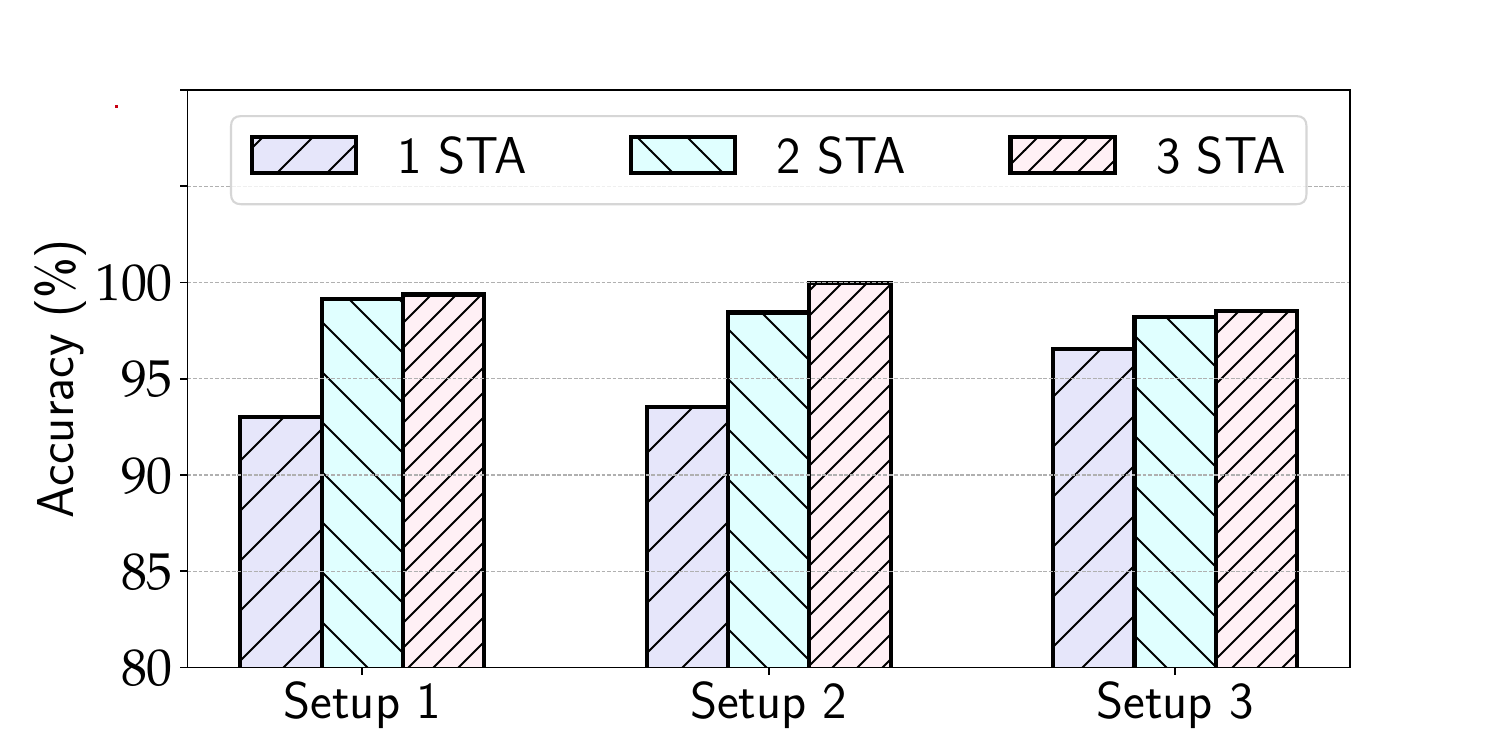}
        \setlength\abovecaptionskip{-0.05cm}
	\caption{Classification performance with \bfas for different number of \glspl{sta}.\vspace{-0.5cm}}
	\label{fig:combined_sta}
\end{figure}

Figure \ref{fig:combined_sta} presents the classification performance of \bfas with baseline \gls{cnn} when we consider a different number of \glspl{sta}. The average classification accuracies are respectively 94.35\%, 98.59\%, and 99.28\% when we use the data from only one \gls{sta}, two \glspl{sta} combined, and three \glspl{sta} combined. It is noticeable that the performance is comparatively poor when we have only one \gls{sta}. 

\begin{figure}[t]
	\centering
	\includegraphics[width=.43\textwidth]{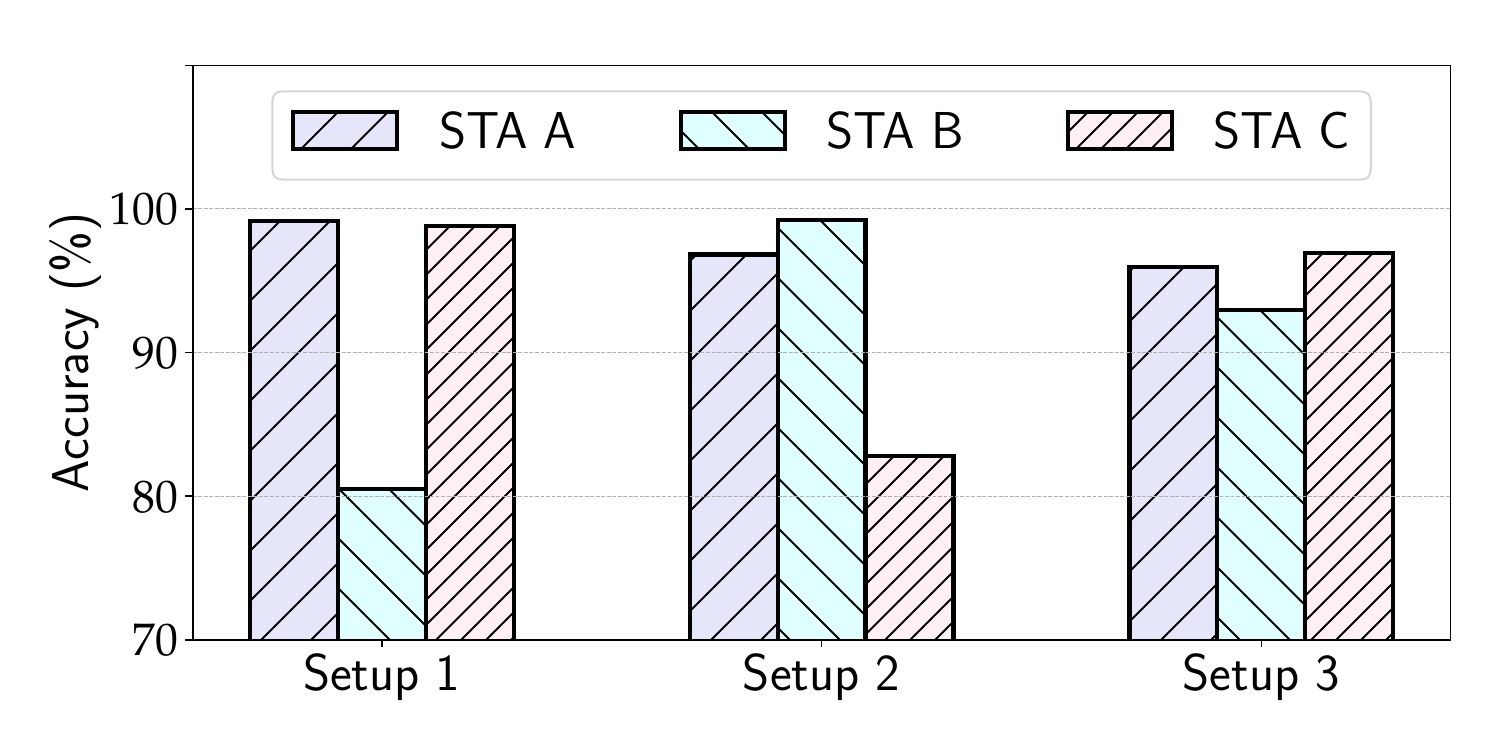}
        \setlength\abovecaptionskip{-0.1cm}
	\caption{Classification performance with \bfas for each individual \gls{sta}.\vspace{-0.1cm}}
	\label{fig:separate_sta}
\end{figure}

To further investigate this aspect, we report the classification performance of each individual beamformees in Figure~\ref{fig:separate_sta}. We notice that, in different setups, the individual performance of the beamformees varies significantly. For example, in setup 1, \gls{sta} B performs worse with an accuracy of around 80\%. On the other hand, in setup 2 and setup 3 \gls{sta} C and \gls{sta} B perform worse with an accuracy of 82.81\% and 92.97\% respectively. This is due to the fact that the physical location of the \gls{sta} may cause the channel between that particular \gls{sta} and the beamformer to be in deep fade, making the model perform poorly. However, this can be improved by considering more than one beamformees for the classification as presented in Figure \ref{fig:combined_sta}.

\begin{figure}[t]
	\centering
	\includegraphics[width=.48\textwidth]{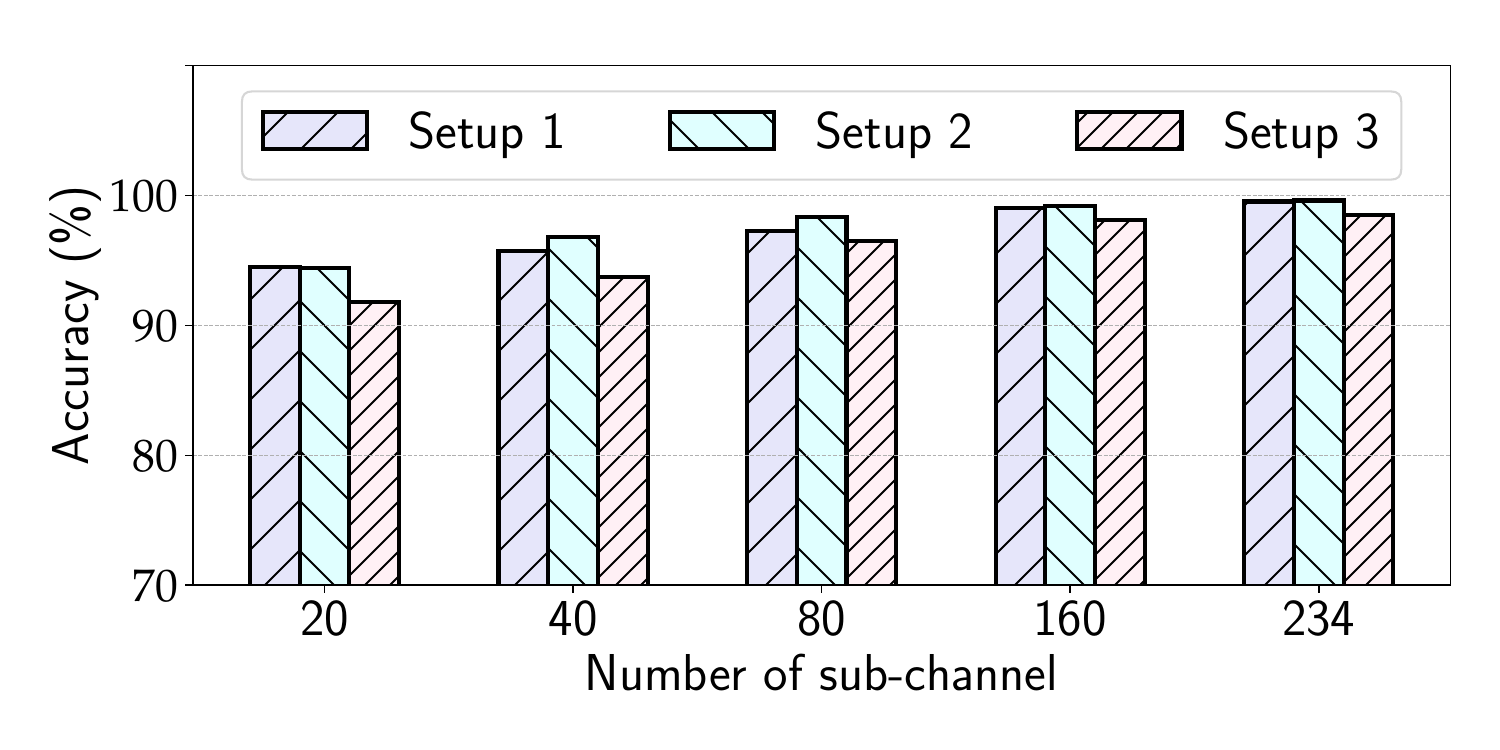}
        \setlength\abovecaptionskip{-0.5cm}
	\caption{Classification performance considering different number of \gls{ofdm} sub-channels\vspace{-0.2cm}}
	\label{fig:different_subcarrier}
\end{figure}
To evaluate the impact of the sub-channel granularity on the activity classification accuracy, we report in Figure \ref{fig:different_subcarrier} the performance of the \gls{cnn} based model, presented in Figure \ref{fig:baseline_cnn}, when we consider different numbers of \gls{ofdm} sub-channels.
The average performance over all the setups considering all 234 data sub-channels at 80~MHz bandwidth is 99.19\%. Even though the overall performance degrades when the number of sub-channels decreases, the reduction in the performance is not significant. On the other hand, a decrease in the number of sub-channels results in a significant reduction of the computational burden. For example, the performance decreases by only 0.44\% and 1.83\% when the number of sub-channels is decreased to 160 and 80, respectively. However, it reduces the computational burden by 1.46 and 2.92 times respectively. Considering only 20 sub-channels, the overall performance is 93.58\% with an 11.7 times reduction of the computational burden. This gives us an idea of the trade-off between the performance and computational burden. 

Overall, the performance analysis shows that \bfas-based sensing performs comparatively with the \gls{cfr} based sensing works of the literature~\cite{ma2019wifi}. \bfas-based sensing can leverage spatial diversity gained through \mum system which most \gls{cfr}-based approaches cannot.\vspace{-0.1cm} 

\section{Related Work}

Over the last few years, a few efforts have been made to leverage \bfas frames for various applications including radio-fingerprinting~\cite{meneghello2022deepcsi} and Wi-Fi sensing~\cite{9700721, kondo2022bi, itahara2022beamforming, haque2023beamsense}. Meneghello et al. leveraged ${\mathbf{\tilde{V}_k}}$ matrices to perform Wi-Fi radio fingerprinting on the move which is based on the intuition that, imperfections in the transmitter's radio circuitry percolate onto the \bfi~\cite{meneghello2022deepcsi}. Kanda et al. extracted ${\mathbf{\tilde{V}_k}}$ matrices from \bfas frames to perform respiratory rate estimation~\cite{9700721}. Kondo et al. evaluate the sensing ability of uni-directional and bi-directional \bfi and demonstrate that \bfi computed from the access point (uplink beamforming) has better sensitivity in comparison to the \bfi computed by the stations (downlink beamforming)~\cite{kondo2022bi}. Itahara et al. leveraged \bfi to estimate the angle of departure for multiple propagation paths and achieved comparable performance to that of \gls{cfr} based approaches~\cite{itahara2022beamforming}. Haque et al. performed sensing by classifying 20 different human activities with compressed \bfas~\cite{haque2023beamsense}. 

Even though there has been increased interest in \bfi-based research, there are no unified open-sourced tools to extract \bfas and reconstruct \gls{bfi} from the captured \bfas. This limits the research community to investigate further this field. Thus we developed \WB to extract \bfas and reconstruct the ${\mathbf{\tilde{V}_k}}$ matrices from any IEEE 802.11 device with all possible bandwidths and network configurations.

\balance
\section{Conclusions and Remarks}
In this paper, we have proposed \WB, the first tool to extract IEEE 802.11ac/ax beamforming feedback angles (BFAs) and reconstruct the beamforming feedback information (BFI) in the wild from both \gls{sum} and \gls{mum} systems operating at 20 MHz/40 MHz/80 MHz/160 MHz bandwidth and with any network configuration. \WB operates at multiple bands and with multiple Wi-Fi standards simultaneously without the need for any direct access to the beamformer or beamformees. We provided an example of leveraging \bfas for one of the most popular wireless sensing tasks, i.e., human activity recognition, achieving up to 99.28\% recognition accuracy. Other \bfi use cases include radio-fingerprinting, Wi-Fi localization, and optimization of the resource allocation in \gls{mum} networks, among others. 
To further promote \bfi-based research we have made the \WB tool compatible with any operating system and we open-sourced it together with a detailed explanation of the operations and practical examples of use.

\begin{acks}
This work is funded in part by the National Science Foundation (NSF) grant CNS-2134973, CNS-2120447, and ECCS-2229472, by the Air Force Office of Scientific Research under contract number FA9550-23-1-0261, by the Office of Naval Research under award number N00014-23-1-2221, and by the Fulbright Schuman Program, administered by the Fulbright Commission in Brussels and jointly financed by the U.S. Department of State, and the Directorate-General for Education, Youth, Sport and Culture (DG.EAC) of the European Commission. The views and opinions expressed in this work are those of the authors and do not necessarily reflect those of the funding institutions.
\end{acks}

\bibliographystyle{unsrt}
\bibliography{reference}

\end{document}